# AI2Pot: A scalable and unified framework for machine-learning interatomic potential development and large-scale molecular dynamic simulations

Hanyu Liu[1], Linggang Zhu[1*], Xuanguang Zhang[1], Ning Yang[1], Jian Zhou[1*], Zhimei Sun[1*]

[1]School of Materials Science and Engineering, Beihang University, Beijing 100191, China.

*Corresponding authors:

lgzhu7@buaa.edu.cn, jzhou@buaa.edu.cn, zmsun@buaa.edu.cn

## Abstract

Machine-learning interatomic potentials (MLIPs) bridge the accuracy of first-principles calculations and the efficiency required for large-scale molecular dynamics (MD) simulations. However, existing MLIP software remains fragmented across different model architectures, making it difficult to establish unified workflows that support flexible model development, efficient training, and scalable MD deployment. Here, we present AI2Pot, a scalable and unified MLIP framework that seamlessly integrates model training, evaluation, and large-scale MD simulations with PyTorch-compatible ecosystem. Instead of relying on generic automatic differentiation for expensive atomistic operators, AI2Pot re-engineers the core computations of Moment tensor potential (MTP) and Neuroevolution potential (NEP) for both training and inference using hand-crafted C++/CUDA code. These specialized operators constitute a unified computational backend shared by training and inference, improving training-inference consistency and reducing memory usage by avoiding large intermediate caches. Benchmarking on the Ge-Sb-Te system demonstrate the accuracy and efficiency of AI2Pot. Deployed via LAMMPS, AI2Pot enables MD of atomic systems containing millions of atoms on a single GPU. Furthermore, AI2Pot provides a companion command-line toolkit (AI2Pot-cli) and Python APIs to facilitate practical MLIP workflows. By unifying high-performance atomistic computing with modern machine-learning ecosystems, AI2Pot offers an user-friendly end-to-end framework for the developing, training, and deploying MLIPs for large scale MD.

## Keywords:

Machine-learning interatomic potentials, Moment tensor potential, Neuroevolution potential, Molecular dynamics

## 1. Introduction

Molecular dynamics (MD) simulations provide a powerful approach for studying the dynamical properties and thermodynamic behavior of materials at the atomic scale. The reliability and efficiency of MD simulations are largely determined by the interatomic potential. *Ab initio* molecular dynamics (AIMD) offers high accuracy by evaluating atomic interactions from electronic-structure calculations, but its computational cost limits the system size and time scales. In contrast, empirical potentials offer significantly improved computational efficiency but typically suffer from limited accuracy and poor transferability across diverse chemical environments. Over the past decades, machine-learning interatomic potentials (MLIPs), trained on first-principles data, have emerged as a compelling alternative that offers a favorable balance between accuracy and computational cost. Representative MLIP architectures include Behle-Parrinello Neural Network potential (BPNN) [1], Gaussian approximation potentials (GAP) [2], Moment Tensor Potential (MTP) [3], Atomic Cluster Expansion potentials (ACE) [4], Deep Potential (DP) [5], and Neuroevolution Potentials (NEP) [6]. These models have been successfully applied to a broad range of materials problems, such as phase transitions [7–9], alloys [10–12], batteries [13–15] and so on. More recently, graph-neural-network-based universal potentials and foundation models, such as M3GNet [16], CHGNet [17], MACE [18], and DPA series models [19], have further expanded the scope of MLIPs by enabling transferable predictions across diverse chemical spaces.

Despite this advances, different MLIP architectures exhibit distinct trade-offs between accuracy and computational cost. GNN-based models usually provide high accuracy through message passing, however, their relatively high inference cost limits their applicability to large-scale molecular dynamics simulations involving complex microstructures, such as grain boundaries, dislocations, and device-scale systems. In contrast, relatively scalable MLIPs, such as MTP, ACE, DP, and NEP models, often adopt more localized descriptors to achieve a favorable balance between accuracy and speed. For example, ACE, DP and NEP have enabled large-scale simulations involving millions to Trillions of atoms [20–23]. These developments highlight the importance of scalable MLIP models.

However, the development of computationally efficient and scalable MLIPs remains fragmented across disparate software ecosystems [21,24–26]. Different potential models are often implemented in independent codes, each with its own input formats, training procedures, and deployment interfaces. This fragmentation imposes a substantial barrier for users seeking to benchmark, compare, or integrate multiple MLIP frameworks within a unified workflow. Furthermore, many MLIP packages rely on self-contained software infrastructures that are only loosely integrated with modern machine-learning ecosystems, such as PyTorch [27] and PyTorch Lightning, where mature components for optimization, scheduling, logging, distributed training, and workflow management are readily available. As a result, current MLIP development and training often lack the flexibility, modularity, and extensibility of standard machine-learning pipelines. These software-level incompatibilities further hinder the construction of unified workflows that seamlessly connect first-principles data generation, MLIP training, model evaluation, and MD deployment.

Here, we present the AI2Pot (Ab Initio & Artificial Intelligence Potential), a scalable and unified MLIP framework that connects model training, evaluation, and deployment for MD simulations. AI2Pot provides unified support for the widely used

MTP and NEP models through custom C++/CUDA operators and PyTorch-compatible training workflows. As a proof of concept, we demonstrate the framework using the example of Ge-Sb-Te dataset and systematically evaluate its GPU-accelerated training efficiency and molecular dynamics inference performance.

# 2. Theory

## 2.1. General formulation of machine-learning interatomic potentials

AI2Pot provides a unified formulation of MLIPs, enabling different model types to be implemented within a consistent computational framework for energy, force, and virial evaluation. Consistent with the conventional formulation of MLIPs, in AI2Pot, the total energy of a system is decomposed into a sum of atomic contributions,

$$E_{tot} = \sum_{i}^{N} E_i \tag{1}$$

where $E_i$ denotes the atomic energy of atom i, determined by its local environment within a cutoff radius.

The local atomic environment of atom i is represented by a set of descriptors $\{D_{i,k}\}$, in which k denotes the descriptor index. The atomic energy is expressed in a unified functional form,

$$E_i = f_\theta(\{D_{i,k}\}) \tag{2}$$

where $f_\theta$ denotes a general learnable mapping. The specific form of $f_\theta$ depends on the selected MLIP architecture. For the Moment Tensor Potential (MTP), $f_\theta$ is formulated as a linear expansion over the descriptors, whereas for the NeuroEvolution Potential (NEP), it is represented by a nonlinear neural network. Under this formulation, different MLIP architectures share the computational scheme from descriptors to energy, while differing only in the learnable mapping function $f_\theta$.

The atomic forces are obtained as the negative gradient of the total energy with respect to atomic positions. By applying the chain rule, the force on atom i along Cartesian direction $\alpha$ can be written as

$$F_{i,\alpha} = \sum_{j\neq i} (\frac{\partial E_i}{\partial r_{ij,\alpha}} - \frac{\partial E_j}{\partial r_{ji,\alpha}}) \tag{3}$$

where $r_{ij} = r_j - r_i$ is the relative position vector between atoms i and j, and the summation runs over all atoms ($j \neq i$) within the cutoff radius.

The virial tensor components $V_{\alpha,\beta}$, where α and β denote Cartesian directions, are computed consistently from atomic energy derivatives as

$$V_{\alpha,\beta} = -\sum_i^N \sum_{j\neq i} \frac{\partial E_i}{\partial r_{ij,\alpha}} r_{ij,\beta} \tag{4}$$

**2.2. Moment tensor Potential**

The MTP implementation in AI2Pot follows the original formulation proposed by Shapeev and the subsequent MLIP-v3 implementation [3,25], with modifications to the radial switching function and other implementation details to improve the performance and integration within the AI2Pot framework. The MTP represents the atomic energy as a linear model over a set of systematically constructed basis functions. The model complexity is controlled by the MTP level, which determines the number of basis functions according to

$$levMTP = 2 + 4\mu + \upsilon \tag{5}$$

where μ denotes the radial basis index. υ denotes the order of the Cartesian tensor product $r_{ij}^{\otimes \upsilon}$, which encodes the angular dependence of the descriptor.

For each atom i, the moment tensor descriptors are constructed by aggregating contributions from neighboring atoms within a cutoff radius. The radial cutoff is defined by two parameters $r_{max}$ and $r_{min}$, where $r_{max}$ is the cutoff radius beyond which atomic interactions vanish, and $r_{min}$ marks the beginning of the smooth switching region. The descriptor components are defined as

$$M_{\mu,\upsilon} = \sum_{j\neq i} \sum_{\zeta}^{N_{basis}} c_{Z_i, Z_j, \mu, \zeta} Q_\zeta (r_{ij}) \tag{6}$$

where $Z_i$ and $Z_j$ denote atomic species of the central and neighboring atoms, respectively. The index $\zeta$ corresponds to the order of the Chebyshev polynomial basis. $N_{basis}$ is the number of Chebyshev basis functions.

The $Q_\zeta (r_{ij})$ is defined as

$$Q_\zeta(r_{ij}) = T_\zeta(x)\, f_c^{MTP}(u) \tag{7}$$

where $T_\zeta$ is the Chebyshev polynomial of order $\zeta$, and the normalized distance x is given by

$$x = \frac{2|r_{ij}| - (r_{max}+r_{min})}{r_{max}-r_{min}} \tag{8}$$

The switching function $f_c^{MTP}(u)$ in the MTP implementation of AI2Pot is designed to improve numerical smoothness at the cutoff boundary. Compared to the original implementation of MTP as in MLIP-v3, the present formulation enforces not only continuity but also vanishing first and second derivatives at the cutoff radius, thereby improving the stability of force and virial evaluations. The switching function $f_c^{MTP}(u)$ is defined as

$$f_c^{MTP}(u) = \begin{cases} 0, & |r_{ij}| \le r_{min} \\ -6u^5 + 15u^4 - 10u^3, & r_{min} < |r_{ij}| \le r_{max} \\ 0 & |r_{ij}| > r_{max} \end{cases} \tag{9}$$

with the normalized distance

$$u = \frac{|r_{ij}| - r_{min}}{r_{max}-r_{min}} \tag{10}$$

The moment tensor descriptors $M_{\mu,\upsilon}$ are further contracted through a predefined tensor contraction scheme to construct scalar invariant basis functions. These basis functions constitute the MTP representation space and are used in the linear expansion of the atomic energy. In AI2Pot, the contraction rules follow the formulation used in MLIP-v3.

### 2.3. Neuroevolution Potential

The NEP re-implemented in AI2Pot is based on the original formulation developed by Fan [6]. In the NEP model, the atomic energies are predicted by a single-hidden-layer neural network operating on a set of rotationally invariant descriptors, where the number of hidden neurons is set to $N_{neurons}$. The descriptor set includes radial (two-body) descriptors $q_n^i$ and angular (many-body) descriptors $q_{nl}^i$, where n indexes the radial basis functions and $l$ and $m$ denote the degree and order associated with spherical harmonic expansions. The numbers of radial and angular basis functions are specified by $n_{max}^{radial}$ and $n_{max}^{angular}$, respectively. As a neural-network-based model, completeness of the descriptor set is not a strict requirement in NEP, since the nonlinear neural network can effectively learn complex atomic interactions from a compact descriptor representation. In the current implementation of AI2Pot, the angular expansion is truncated at $l_{max}$=4 and descriptor set includes only two-body and three-body terms.

The radial descriptors are defined as

$$q_n^i = \sum_{j\neq i} \sum_{\zeta}^{N_{basis}^{radial}} c_{Z_i,Z_j,n,\zeta}^{radial} Q_\zeta(r_{ij}) \quad (11)$$

where $Z_i$ and $Z_j$ denote atomic species of the central and neighboring atoms, respectively. $N_{basis}^{radial}$ is the number of Chebyshev basis functions for radial descriptors. $Q_\zeta(r_{ij})$ is constructed as

$$Q_\zeta(r_{ij}) = \frac{T_\zeta(x)+1}{2} f_c^{NEP}(r_{ij}) \quad (12)$$

where $T_\zeta(x)$ is the Chebyshev polynomial of order $\zeta$, and the normalized distance is given by

$$x = 2(\frac{|r_{ij}|}{r_{max}} - 1)^2 - 1 \quad (13)$$

The switching function in NEP is defined as

$$f_c^{NEP} = \begin{cases} \frac{1+\cos u}{2}, & |r_{ij}| \leq r_{max} \\ 0, & |r_{ij}| > r_{max} \end{cases} \quad (14)$$

with the normalized distance

$$u = \pi \frac{|r_{ij}|}{r_{max}} \quad (15)$$

The angular descriptors $q_{nl}^i$ are constructed based on the addition theorem of spherical harmonics. An intermediate quantity is defined as

$$S_{nlm}^i = \sum_{j\neq i} \sum_{\zeta}^{N_{basis}^{angular}} c_{Z_i,Z_j,n,\zeta}^{angular} Q_\zeta(r_{ij}) \frac{b_{lm}(x_{ij},y_{ij},z_{ij})}{r^l} \quad (16)$$

where $b_{lm}(x_{ij}, y_{ij}, z_{ij})$ denotes real spherical harmonic basis functions.

$N_{basis}^{angular}$ is the number of Chebyshev basis functions for angular descriptors. The final 3-body descriptor is obtained as

$$q_{nl}^{i} = \sum_{m=0}^{2l} C_{lm} (S_{nlm}^{i})^2 \quad (17)$$

where $C_{lm}$ are constants obtained from the addition theorem of spherical harmonics.

In AI2Pot, the construction of $S_{nlm}^{i}$ follows the same computational pattern as the moment tensor descriptors in MTP, ensuring a unified descriptor backend across different MLIP models within the framework.

### 2.4. Loss function

The loss function for potential model training in AI2Pot is defined at the configuration level, combining energy, force, and virial contributions into a unified objective. The total loss is given by

$$L_{total} = L_E + L_F + L_V \quad (18)$$

The energy, force, and virial loss terms are defined as follows:

$$L_E = \frac{w_E}{N} (E^{MLIP} - E^{DFT})^2 \quad (19)$$

$$L_F = \frac{w_F}{3N} \sum_{i}^{N} \sum_{\alpha}^{3} (F_{i,\alpha}^{MLIP} - F_{i,\alpha}^{DFT})^2 \quad (20)$$

$$L_V = \frac{w_V}{9N} \sum_{\alpha}^{3} \sum_{\beta}^{3} (V_{\alpha,\beta}^{MLIP} - V_{\alpha,\beta}^{DFT})^2 \quad (21)$$

where $w_E$, $w_F$, and $w_V$ are weighting factors that balance the contributions of energy, force, and virial terms.

## 3. AI2Pot Framework and Implementation

### 3.1. Architecture

As illustrated in **Figure 1**, AI2Pot adopts a layered architecture that separates high-level MLIP models development from performance-critical atomistic computation. The top layer in **Figure 1** is a PyTorch/Lightning frontend, which provides user-facing components for model definition, training, checkpointing, and evaluation. This layer preserves the flexibility of the PyTorch ecosystem and allows AI2Pot to integrate naturally with modern machine-learning workflows. The core computational layer in **Figure 1** includes custom C++/CUDA operators. These operators are responsible for descriptor construction, energy evaluation, force calculation, virial calculation, and training-related operations. A key design principle of AI2Pot is that the same operator backend is shared by both training and inference. This unified backend avoids discrepancies between model optimization and deployment, improves training-inference consistency, and reduces the engineering overhead of maintaining separate implementations. The connection between the frontend and the custom backend in **Figure 1** is realized through Torch bindings. This interface enables the custom operators to participate directly in the PyTorch training pipeline while retaining execution efficiency of custom C++/CUDA kernels for performance-critical computations. In this way, AI2Pot combines the programmability of PyTorch with the computational efficiency of dedicated C++/CUDA kernels. The bottom layer in **Figure 1** provides inference and deployment interfaces for molecular dynamics applications. Trained models can be deployed to external simulation engine such as LAMMPS and ASE, allowing AI2Pot to support atomistic simulations beyond model development. Overall, this architecture establishes a complete route from model training to high-performance inference and MD deployment.

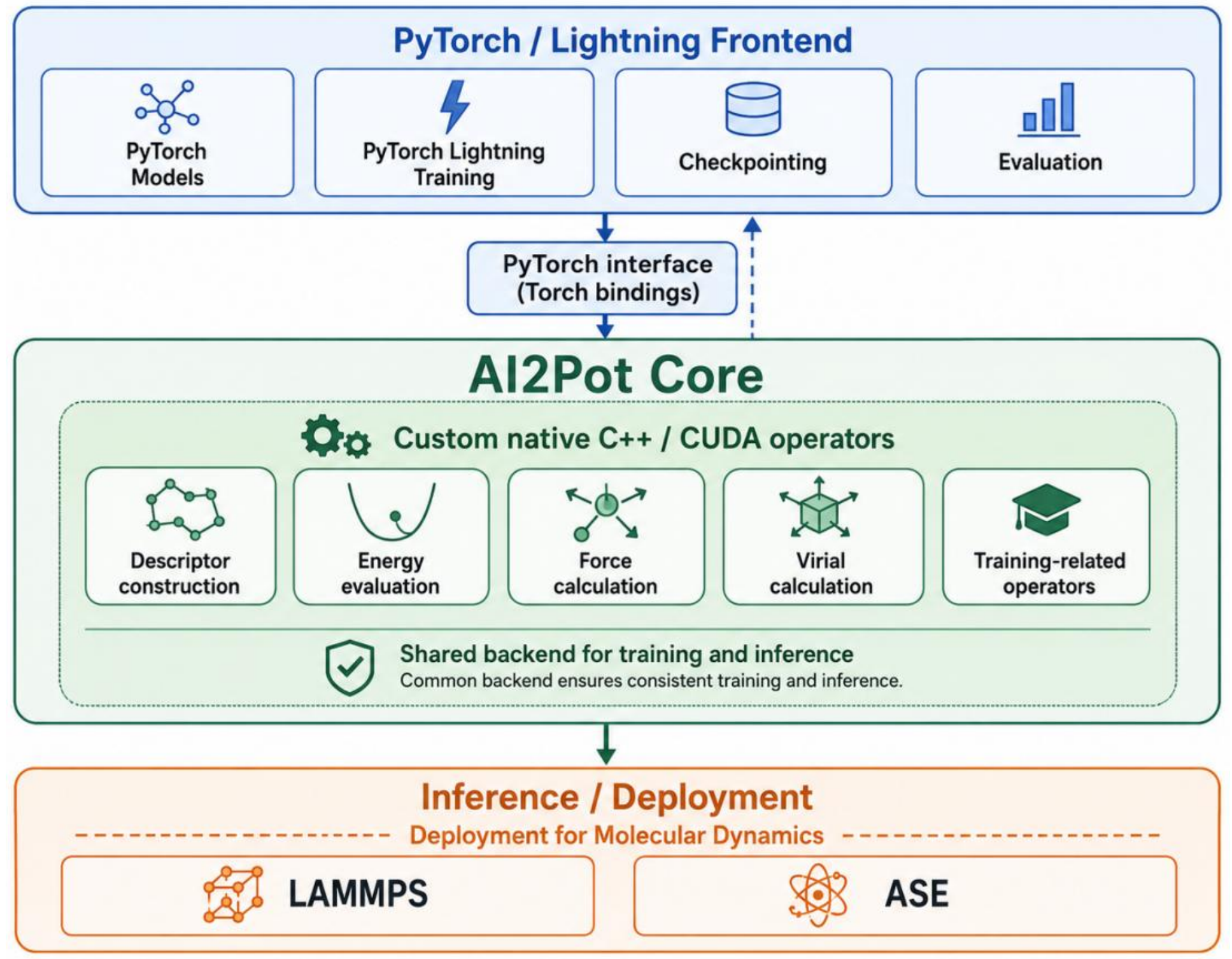

**Figure 1.** Architecture of the AI2Pot, consisting of a PyTorch/Lightning frontend, a custom C++/CUDA computational backend, and molecular dynamics deployment interfaces.

### 3.2. Training and inference with shared C++/CUDA operators

The computational design of core C++/CUDA operators in AI2Pot follows a per-atom execution paradigm, as illustrated in **Figure 2**. In contrast to conventional tensor-parallel strategies widely used in deep learning frameworks, where workloads are distributed over dense tensor blocks, AI2Pot assigns independent computation to each central atom. This atom-wise computation pattern directly maps to the design of low-level C++/CUDA kernels in AI2Pot. Within this framework, all core operators are implemented as unified C++/CUDA kernels shared between training and inference, ensuring numerical consistency across different execution stages. The same operators are reused for both model optimization and molecular dynamics inference, avoiding redundant implementations of forward and backward computations. All operators in AI2Pot support both float32 and float64 precision. Their correctness has been validated using PyTorch autograd gradient checking via

`torch.autograd.gradcheck()`.

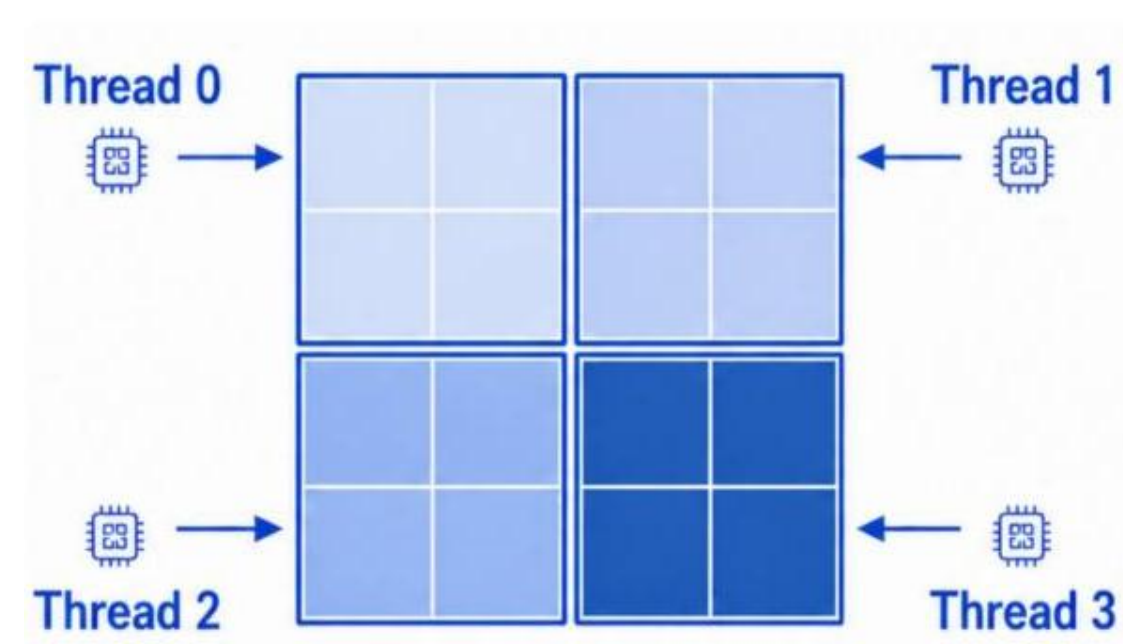

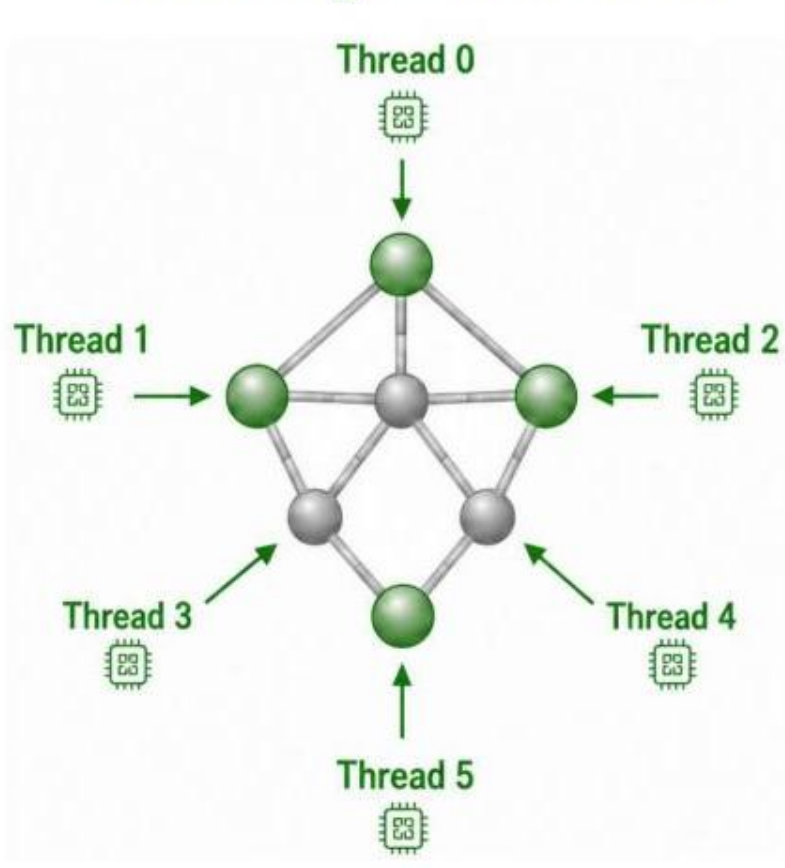

**Figure 2.** Comparison between tensor-parallel execution in PyTorch and atom-parallel execution in AI2Pot operators.

The resulting computational efficiency is demonstrated in **Figure 3**, which reports the performance scaling of the core training operators as a function of system size. In this demonstration, we focus on the training operators for MTP and NEP models, implemented as `ai2pot.fromcc.linearMtpToLossOp` and `ai2pot.fromcc.nepToLossOp`, respectively. The performance of inference operators is discussed separately in **Section 4.3**. A cubic $Ge_2Sb_2Te_5$ supercell with random atomic occupations is used as the benchmark system. Since typical training datasets rarely contain large structures, the system size is varied up to 900 atoms to evaluate scaling behavior under realistic training conditions. For the MTP training operator, OpenMP parallelization with 16 and 32 threads achieves a speedup of approximately 3-6× relative to single-thread execution, and remains nearly constant across system sizes. In contrast, GPU acceleration provides a speedup close to 10× at 99 atoms, and continues to increase with system size. For the NEP training operator, OpenMP parallelization achieves a stronger speedup of approximately 5-8× under 16 and 32 threads, also exhibiting weak dependence on system size. GPU acceleration further improves performance, reaching a speedup of approximately 14× at 99 atoms, and increasing continuously with system size. Compared to MTP, the NEP training operator demonstrates superior GPU scaling. At 900 atoms, the GPU speedup reaches 125.38× for NEP, compared to 77.96× for MTP. This difference originates from the higher memory overhead of the MTP moment tensor descriptors $M_{\mu,\upsilon}$, which are reallocated during each forward and backward pass, whereas the NEP intermediate quantities $S^{i}_{nlm}$ require significantly lower memory consumption.

In summary, GPU execution provides a favorable scaling regime for training large supercells, while also maintaining competitive performance at small system sizes. In this benchmark, the GPU is not fully saturated, suggesting that faster training can be achieved by increasing batch size.

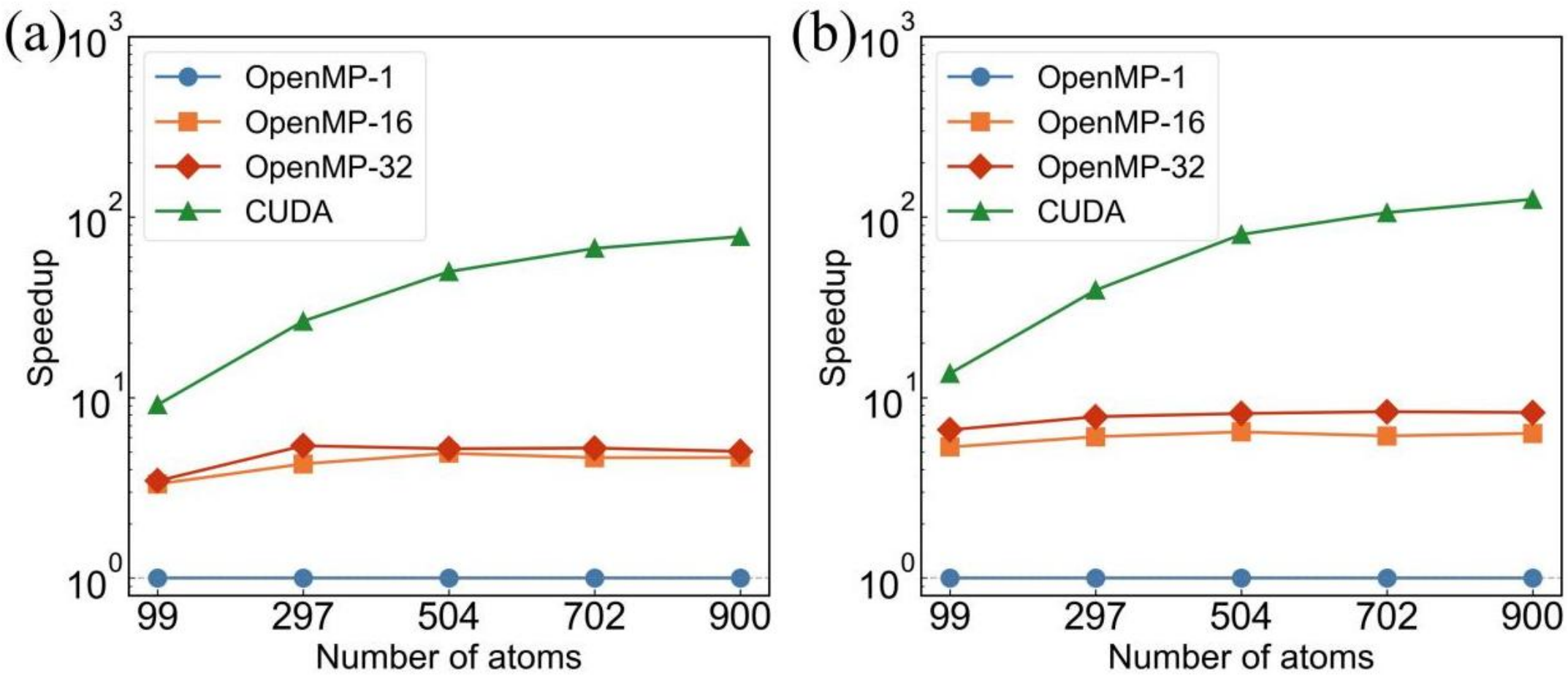

**Figure 3.** Scaling behavior of MTP and NEP training operators on CPU (OpenMP) and GPU (CUDA) with increasing system size.

### 3.3. Optimization protocol

The training pipeline of MLIPs in AI2Pot is implemented using the PyTorch Lightning framework, which provides a flexible and modular interface for training control. This design allows optimizer configuration, learning-rate scheduling, dynamic loss weighting, gradient clipping, and callback-based operations to be implemented in a unified and reproducible manner.

#### 3.3.1. Optimizer configuration

The optimization strategy in AI2Pot differs from conventional workflows used in several representative MLIP implementations. In the original MTP framework, the model is commonly optimized by combining BFGS-type nonlinear optimization with linear least-squares updates for the linear coefficients [25]. Subsequent MTP implementations have also employed L-BFGS-based optimization together with least-squares solvers [28,29]. In contrast, AI2Pot treats all parameters of MTP models as learnable tensors and optimizes them directly using a first-order gradient-based optimizer within the PyTorch training framework. For NEP models, the original GPUMD implementation employs neuroevolution-based optimization [21], whereas later developments have introduced gradient-based optimizers such as AdamW [30,31]. AI2Pot follows this gradient-based direction and provides a unified optimizer interface for both MTP and NEP models. This design allows different potential models to be trained under the same optimization framework.

In the examples shown in this work, the model parameters are optimized using the Adam optimizer. The optimizer is applied to all learnable parameters of each MLIP model, with an initial learning rate of $lr_{start}$. The exponential moving-average coefficients are set to $\beta_1$ = 0.9 and $\beta_2$ = 0.999, and the numerical-stability constant is fixed to $\varepsilon = 10^{-8}$. No weight decay is applied during training. This configuration provides a stable and robust optimization framework for both linear MTP and neural-network-based NEP models, ensuring efficient convergence across different model complexities.

### 3.3.2. Learning-rate scheduling

AI2Pot employs a two-stage learning-rate schedule consisting of a linear warmup stage followed by cosine annealing decay. The learning rate is updated at every training step, and the schedule is parameterized by an initial learning rate $lr^{start}$ and a final learning rate $lr_{end}$. The warmup stage is introduced to stabilize early-stage optimization and prevent abrupt gradient updates. It is implemented using `LinearLR` scheduler in PyTorch, where the learning rate is increased linearly from a fixed start factor of 1e-4 to $lr^{start}$. The warmup duration is set to one epoch. After warmup, the learning rate follows a cosine annealing schedule implemented via `CosineAnnealingLR` in PyTorch, gradually decaying from $lr^{start}$ to $lr^{end}$ over the remaining training steps. This design ensures a smooth transition from rapid initial adaptation to stable fine-grained convergence. The entire learning-rate scheduling pipeline is implemented by overriding the `configure_optimizers()` method in PyTorch Lightning's `LightningModule` class.

### 3.3.3. Dynamic loss weighting

AI2Pot employs a dynamic loss weighting strategy to balance the contributions of energy, force, and virial terms during training. The weighting factors are updated at every training step according to a linear interpolation scheme between predefined initial and final values. The weights of energy, force, virial loss are defined as

$$w_e^{\tau} = w_e^{start} \times lr^{\tau} + w_e^{end} \times (1 - lr^{\tau})$$

$$w_f^{\tau} = w_f^{start} \times lr^{\tau} + w_f^{end} \times (1 - lr^{\tau})$$

$$w_v^{\tau} = w_v^{start} \times lr^{\tau} + w_v^{end} \times (1 - lr^{\tau})$$

where $w_e^{\tau}$, $w_f^{\tau}$ and $w_v^{\tau}$ denote the loss weights of energy, force and virial at training step $\tau$, respectively. And $lr^{\tau}$ is the learning rate at the same step. $w_c^{\tau}$ $(c \in \{e, f, v\})$ are the predefined initial and final weights for each loss component.

## 3.4. Training execution workflow

The training execution workflow of AI2Pot is illustrated in **Figure 4**. AI2Pot adopts an on-the-fly neighbor-list construction strategy without caching, closely following the pattern of MD simulations. This design ensures consistency between training and inference. As shown in **Figure 4**, training is initialized by loading atomic structures from Extxyz files using the ASE interface. At each training step, atomic structures are first parsed and passed to the AI2Pot neighbor-list class (`ai2pot.core.Nblist`) for dynamic construction of local atomic environments. The resulting data are then transferred to the target computing device (CPU or GPU), where forward evaluation of energies, forces, and virials is performed using AI2Pot operators, followed by backpropagation for parameter optimization. Due to the memory efficiency of the on-the-fly neighbor-list construction and operator-based computation, the current implementation only supports training on CPU or single-GPU.

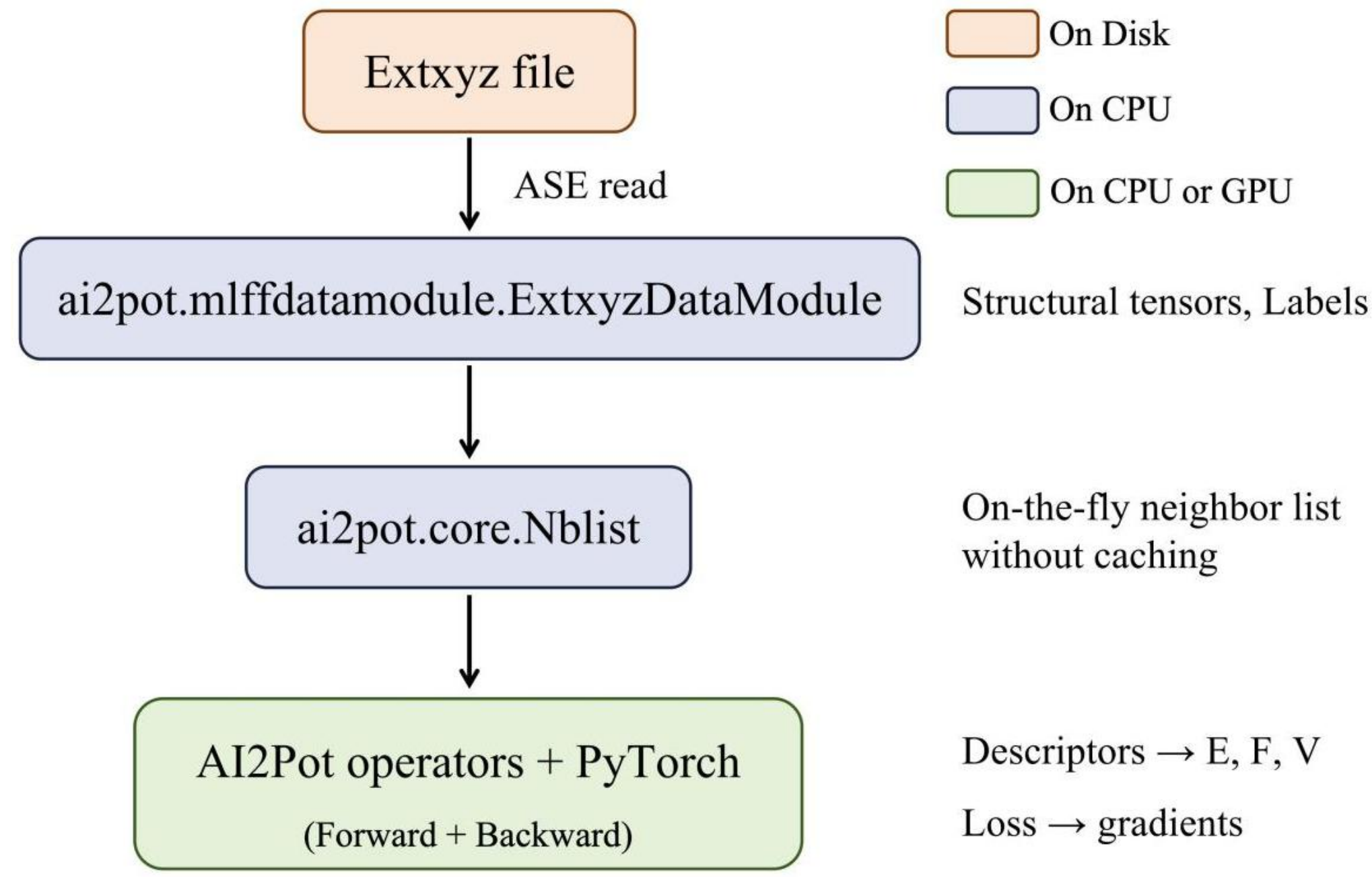

**Figure 4.** Training execution workflow of AI2Pot, including on-the-fly neighbor list construction and device-aware execution.

This Training execution workflow is exposed through a Python interface, allowing users to construct datasets, initialize models, and launch training in a manner similar to using a standard Python package. For example, a MTP model with default hyperparameters can be trained using the following minimal script:

```python
from lightning import Trainer
from ai2pot.data.mlffdataset import ExtxyzDataset
from ai2pot.data.mlffdatamodule import ExtxyzDatamodule
from ai2pot.models.potential_train import LitLinearMtp

type_map = ExtxyzDataset.get_type_map(trainset_path)
trainer = Trainer(max_epochs)
datamodule = ExtxyzDatamodule(trainset_path, validset_path,
batch_size)
```

```
lit_linear_mtp = LitLinearMtp(type_map=type_map)
trainer.fit(model=lit_linear_mtp, datamodule=datamodule)
```

In this script, the `datamodule` handles structure loading, batching, neighbor-list preparation, and data transfer, whereas the `lit_linear_mtp` encapsulates the potential model, loss evaluation, optimizer definition, and training-step logic.

**3.5. Model export and deployment in ASE/LAMMPS**

AI2Pot provides a unified model export and deployment framework that enables consistent inference across ASE and LAMMPS, as shown in **Figure 5**. The same trained model parameters are shared across all deployment backends, while different serialization formats are used for ASE and LAMMPS execution. For ASE, AI2Pot models are deployed using PyTorch Lightning checkpoints. This format preserves full model flexibility, including dynamic computation graphs and configurable precision. In contrast, LAMMPS deployment relies on TorchScript-compiled models (`torch.jit.script()`) serialized as static `.pt` files. To ensure execution stability and compatibility with the C++ runtime, LAMMPS inference is restricted to single-precision floating-point. Although this introduces reduced flexibility compared to checkpoint-based execution, it significantly improves robustness in large-scale distributed simulations.

For ASE-based simulations, AI2Pot provides Python-compatible calculators for both MTP and NEP models. The usage is straightforward through the ASE calculator interface:

```python
# MTP calculator
from ai2pot.models.mtp.linear_mtp_utils import LinearMtpCalculator
mtp_calculator = LinearMtpCalculator(checkpoint_path, map_location, torch_float_dtype)
atoms = ase_read(POSCAR_path)
atoms.calc = mtp_calculator

# NEP calculator
from ai2pot.models.nep.nep_utils import NepCalculator
```

---

```
nep_calculator = NepCalculator(checkpoint_path, map_location, torch_float_dtype)
atoms = ase_read(POSCAR_path)
atoms.calc = nep_calculator
```

During ASE runtime, atomic structures are processed on the CPU, where neighbor lists are constructed using `ai2pot.core.Nblist` with linear complexity. The resulting neighbor information is transferred to the target device (CPU or GPU) for evaluation of energies, forces, and virials. The computed energies are returned as scalars, while forces and virials are converted to NumPy arrays for downstream processing.

For LAMMPS integration, AI2Pot is invoked via the standard pair style interface as follows:

```txt
pair_style ai2pot libtorch_mlip.pt
pair * * Ge Sb Te
```

This command in the LAMMPS input file initializes the AI2Pot potential and triggers the evaluation of atomic interactions during MD. AI2Pot adopts an MPI-parallel execution model in which each process is mapped to a computional device (CPU or GPU). Neighbor lists are constructed on the CPU for each local domain and converted into Torch tensors before GPU inference. The computed energies, forces, and virials are then communicated back to the LAMMPS engine for large-scale MD simulations.

Overall, AI2Pot uses the same trained model and computational operators across all deployment backends, while adapting only the serialization format to different simulation environments.

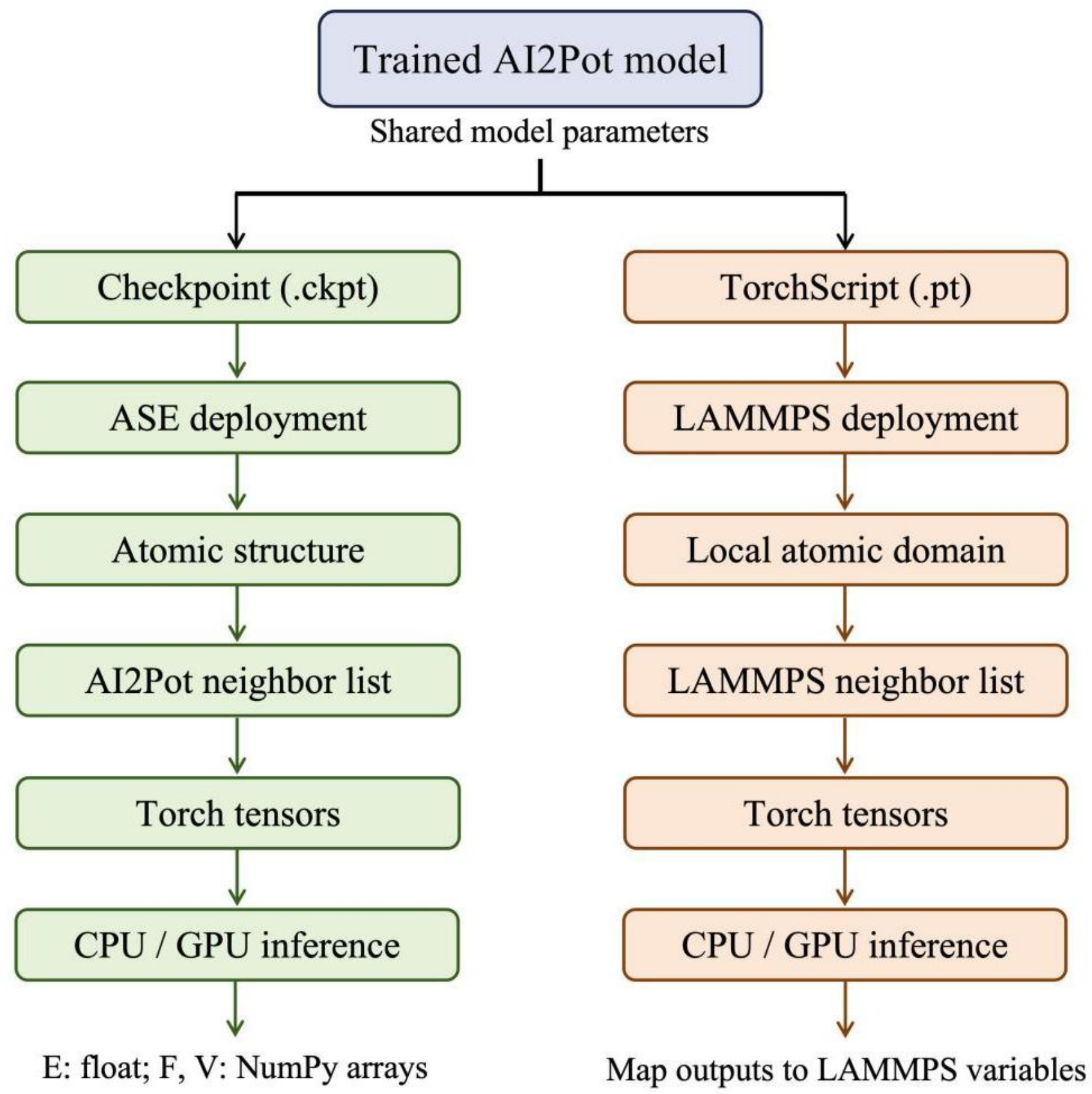

**Figure 5.** Deployment workflow of AI2Pot across ASE and LAMMPS backends.

### 3.6. AI2Pot-cli

AI2Pot provides a Python-based interface integrated into a broad computational ecosystem for accessing high-performance atomistic simulation and machine-learning routines, while AI2Pot-cli is built on top of this interface as a higher-level command-line layer that enables standardized and low-configuration workflow execution. AI2Pot-cli is inspired by workflow-oriented interfaces such as VASPKIT and serves as a unified entry point for both interactive and script-based execution. As shown in **Figure 6**, the AI2Pot-cli interface begins with a banner section that provides essential metadata, including the software name, version information, developer identity, and repository links for both AI2Pot and AI2Pot-cli. This design ensures traceability, version control awareness, and reproducibility across different computational environments. The main interface is organized into modular functional groups, including installation, data preprocessing, training input generation, post-processing, and molecular dynamics utilities. Users can access an interactive menu-driven mode by executing `ai2pot-cli`, which enables common operations such as dataset conversion, descriptor analysis, and visualization of model performance metrics.

In addition to interactive usage, AI2Pot-cli supports fully scriptable execution. For example, an MTP model can be trained with only two command-line steps:

```shell
$ echo -e 21 | ai2pot-cli
$ ai2pot-cli train --input mtp_train.jsonc
```

It should be noted that the current version of AI2Pot-cli does not yet include active learning functionality and MD utilities. These features are planned for future development.

Overall, AI2Pot-cli provides a lightweight and convenient workflow interface that bridges data preparation, model construction, evaluation, and deployment within the AI2Pot framework.

```
(ai2pot_env) [liuhanyu@localhost ~]$ ai2pot-cli
+------------------------------------------------------------------------------+
|                          AI2Pot-cli Standard Edition                         |
|                               Version 1.0.0                                  |
|                                                                              |
|                  Official Command Line Interface for AI2Pot                  |
|                                                                              |
|                Developer: Hanyu Liu (hyliu2016@buaa.edu.cn)                  |
|                                                                              |
|             AI2Pot-cli : https://github.com/lhycms/AI2Pot-cli                |
|                 AI2Pot : https://github.com/lhycms/AI2Pot                    |
+------------------------------------------------------------------------------+

=============================== Installation ===============================
 1)  Install AI2Pot                   2)  Install LAMMPS with AI2Pot
============================== Preprocessing ===============================
11)  Convert Dataset                 12)  Standardize ExtXYZ
13)  Analyse Dataset                 14)  MTP Active Learning
15)  NEP Active Learning
========================= Potential Training Input =========================
21)  MTP Training Input              22)  NEP Training Input
============================== Postprocessing ==============================
31)  Plot E/F/V Parity               32)  Plot Learning Curve
33)  Plot Descriptor Projection      34)  Export TorchScript Model
=============================== MD Utilities ===============================
91)  Doctor                          92)  Show Examples
93)  Print Version
 0)  Quit
------------>>
```

**Figure 6.** Main interface of AI2Pot-cli.

## 4. Application and Performance Evaluation

### 4.1. Potential training for Ge-Sb-Te ternary system

This section demonstrates the training capability of AI2Pot using a Ge-Sb-Te dataset as a representative example. It should be noted that the accuracy of MLIP models depends strongly on the choice of model hyperparameters and training settings. In general, longer training, smaller batch sizes, and more carefully optimized hyperparameters may further improve the final model accuracy. Therefore, the results presented in this section should not be interpreted as the upper limit of the MTP or NEP model performance. Instead, they just represent the performance obtained using the default training configurations provided in AI2Pot and AI2Pot-cli.

The Ge-Sb-Te dataset used in this study is divided into a training set and a test set. The training set contains 2633 frames with 283465 atoms in total, while the test set contains 673 frames with 85821 atoms in total. The dataset covers crystalline, liquid, and amorphous configurations with different chemical compositions in the Ge-Sb-Te system. First-principles energies, atomic forces, and virial tensors are provided for these structures and used as reference labels for supervised MLIP training and evaluation. All benchmarks in this section were performed on a workstation equipped with an Intel Xeon Platinum 8352Y CPU @ 2.20 GHz and an NVIDIA GeForce RTX 4090 D GPU.

#### 4.1.1. MTP training

**Table 1** summarizes the hyperparameters used for MTP training. Except for the batch size, which is set to 1 in this study, all other parameters follow the default MTP training configuration generated by AI2Pot-cli. The combination of MTP level 18 and three chemical elements in the Ge-Sb-Te dataset results in 526 trainable parameters, corresponding to a MLIP model with a relatively small parameter count.

**Table 1.** Hyperparameters used for training the MTP model on the Ge-Sb-Te dataset.

| Parameters | Value |
|---|---|
| levMTP | 18 |
| $r_{max}$ | 5.0 |
| $r_{min}$ | 0.0 |
| $N_{basis}$ | 8 |
| $lr_{start}$ | 1e-02 |
| $lr_{end}$ | 1e-04 |
| $w_e^{start}$ | 0.1 |
| $w_e^{end}$ | 2.0 |
| $w_f^{start}$ | 10.0 |
| $w_v^{end}$ | 1.0 |
| $w_v^{start}$ | 0.1 |
| $w_v^{end}$ | 1.0 |
| *Epochs* | 200 |
| *Batch size* | 1 |

The training behavior and prediction accuracy of the MTP model are summarized in **Figure 7**. **Figure 7(a)** shows the evolution of the energy, force, and virial losses during training. Because dynamic loss weighting is used, as shown in **Figure 7(b)**, the force term has a larger weight at the early stage of training. As a result, the force loss decreases mainly in the early training stage, whereas the energy and virial losses continue to decrease at later stages as their relative weights increase. **Figure 7(c)** presents the principal component analysis (PCA) of the MTP descriptors for the training and test sets. The test structures are largely covered by the descriptor distribution of the training set, indicating that the test set lies within the descriptor-space domain of training set. **Figures 7(d-f)** show the parity plots for energy, force, and virial predictions, respectively. The MTP model achieves training RMSEs of 21.79 meV/atom for energy, 243.86 meV/Å for force, and 74.40 meV/atom for virial. On the test set, the corresponding RMSEs are 25.03 meV/atom, 207.69 meV/Å, and 46.38 meV/atom, respectively. These results demonstrate that the MTP model trained with the default AI2Pot-cli configuration provides reasonable predictive accuracy for the Ge-Sb-Te dataset.

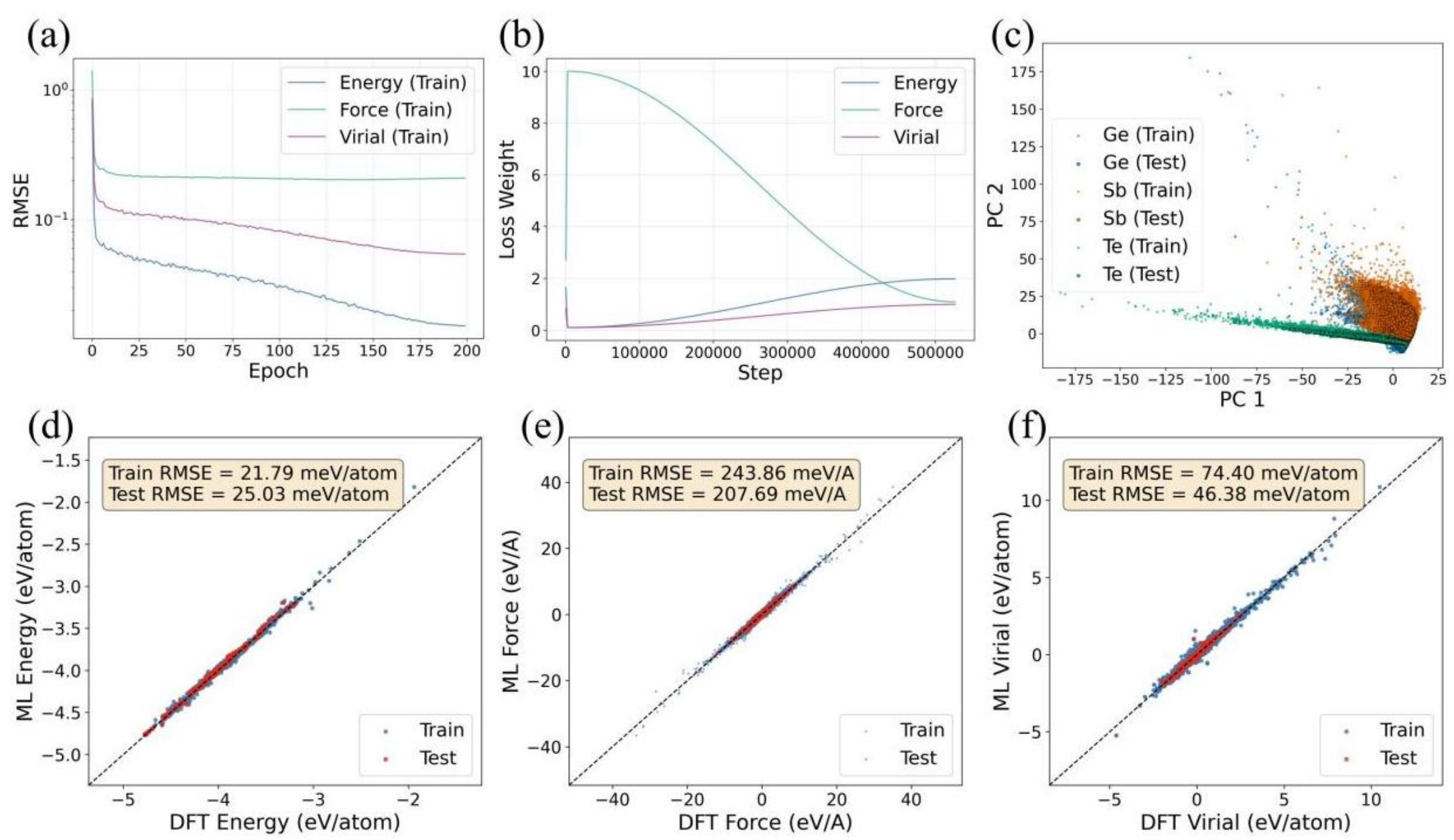

**Figure 7.** Training behavior, descriptor-space distribution, and prediction accuracy of the MTP model on the Ge-Sb-Te dataset. (a) Learning curves showing the evolution of training and validation losses during MTP training. (b) Dynamic loss-weight evolution for the energy, force, and virial terms. (c) PCA of MTP descriptors for the training and test sets. (d-f) RMSEs of energy, force, and virial predictions for the training and test sets, respectively.

As shown in Figure 8, the force RMSE remains nearly unchanged over the entire batch-size range, with the test-set error ratio fluctuating within 0.05 of the batch size=1 baseline. In comparison, the energy and virial errors are more sensitive to increasing batch size. The energy train RMSE increases to 1.3 at batch size = 32. The virial RMSE ratio shows a more monotonic degradation, reaching 1.4 at the largest batch size. These results highlight the trade-off between training efficiency and model accuracy with respect to batch size. Large batch sizes (e.g., 16 or above) are suitable for early-stage iterative dataset construction. In contrast, smaller batch sizes (e.g., 1 - 4) are recommended for final model training when high prediction accuracy is required.

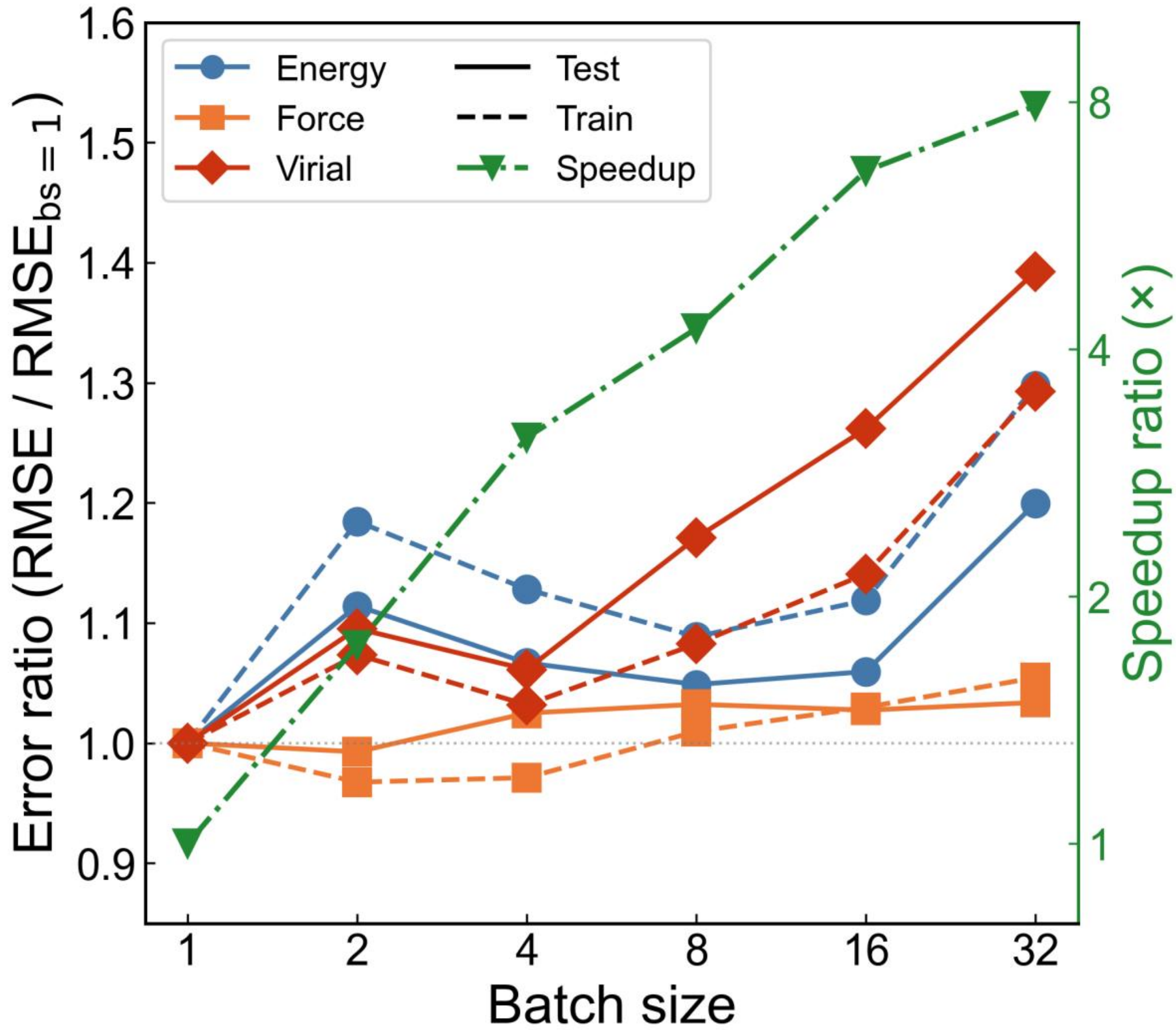

**Figure 8.** Effect of batch size on MTP accuracy and training efficiency.

### 4.1.2. NEP training

**Table 2** summarizes the hyperparameters used for NEP training. Except for the batch size, which is set to 1 in this study, all other parameters follow the default NEP training configuration generated by AI2Pot-cli. This NEP has 7.9k trainable parameters.

**Table 2.** Hyperparameters used for training the MTP model on the Ge-Sb-Te dataset.

| Parameters | Value |
|---|---|
| $n_{max}^{radial}$ | 10 |
| $n_{max}^{angular}$ | 8 |
| $l_{max}$ | 4 |
| $r_{max}^{radial}$ | 7.0 |
| $r_{max}^{angular}$ | 5.0 |
| $N_{basis}^{radial}$ | 8 |
| $N_{basis}^{angular}$ | 8 |
| $N_{neurons}$ | 50 |
| $lr_{start}$ | 1e-03 |
| $lr_{end}$ | 1e-06 |
| $w_{e}^{start}$ | 0.1 |
| $w_{e}^{end}$ | 2.0 |
| $w_{f}^{start}$ | 10.0 |
| $w_{v}^{end}$ | 1.0 |
| $w_{v}^{start}$ | 0.1 |
| $w_{v}^{end}$ | 1.0 |
| *Epochs* | 200 |
| *Batch size* | 1 |

The training behavior and prediction accuracy of the NEP model are summarized in **Figure 9**. Similar to the MTP case, the dynamic loss weighting scheme leads to a larger force weight at the early stage of training, resulting in an early decrease in the force loss, followed by further reduction of the energy and virial losses as their relative weights increase, as shown in **Figures 9(a,b)**. **Figure 9(c)** shows the PCA distribution of NEP descriptors for the training and test sets. The test structures are largely covered by the descriptor distribution of the training set, indicating that the test set remains within the descriptor-space domain sampled during training. **Figures 9(d-f)** present the parity plots for energy, force, and virial predictions, respectively.

The NEP model achieves training RMSEs of 12.74 meV/atom for energy, 176.38 meV/Å for force, and 47.58 meV/atom for virial. On the test set, the corresponding RMSEs are 16.47 meV/atom, 156.28 meV/Å, and 31.22 meV/atom, respectively. These results indicate that the NEP model trained with the default AI2Pot-cli configuration achieves improved predictive accuracy for the Ge-Sb-Te dataset compared with the compact MTP model discussed above.

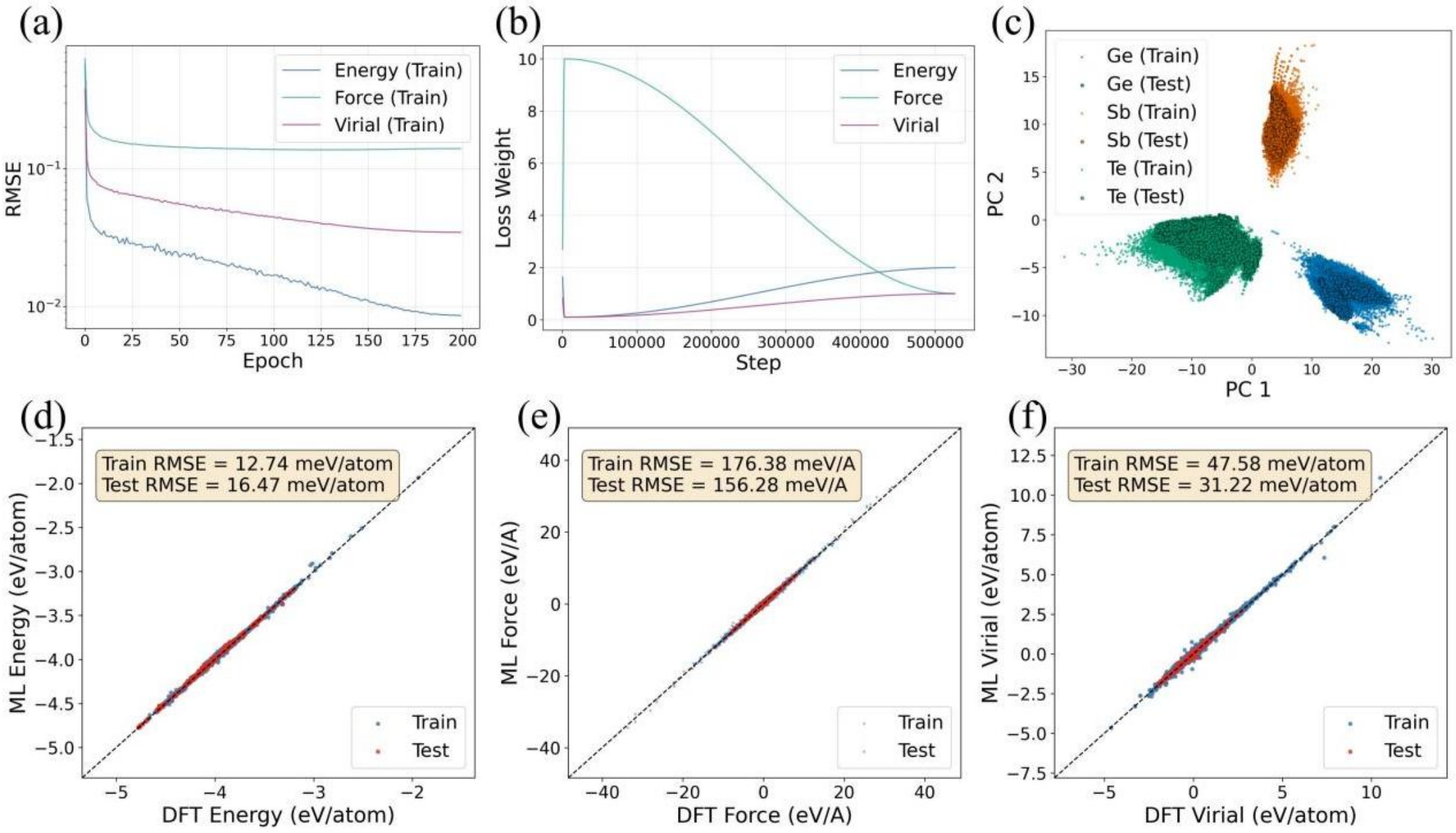

**Figure 9.** Training behavior, descriptor-space distribution, and prediction accuracy of the NEP model on the Ge-Sb-Te dataset. (a) Learning curves showing the evolution of training and validation losses during NEP training. (b) Dynamic loss-weight evolution for the energy, force, and virial terms. (c) PCA of NEP descriptors for the training and test sets. (d-f) RMSEs of energy, force, and virial predictions for the training and test sets, respectively.

For the NEP model, as shown in Figure 10, the force RMSE exhibits a moderate upward drift with increasing batch size, with the test-set error ratio rising to 1.13 at batch size = 32. The energy error proves considerably more sensitive: the train-set RMSE ratio climbs sharply to approximately 1.5, while the test-set ratio displays a non-monotonic dependence, dropping below the baseline at intermediate batch sizes before recovering to near unity. The virial RMSE ratio again degrades monotonically, reaching 1.4 at the largest batch size. Similar to MTP, larger batch sizes can improve training efficiency during rapid exploration of candidate structures, whereas smaller batch sizes are preferred for final model refinement requiring higher accuracy.

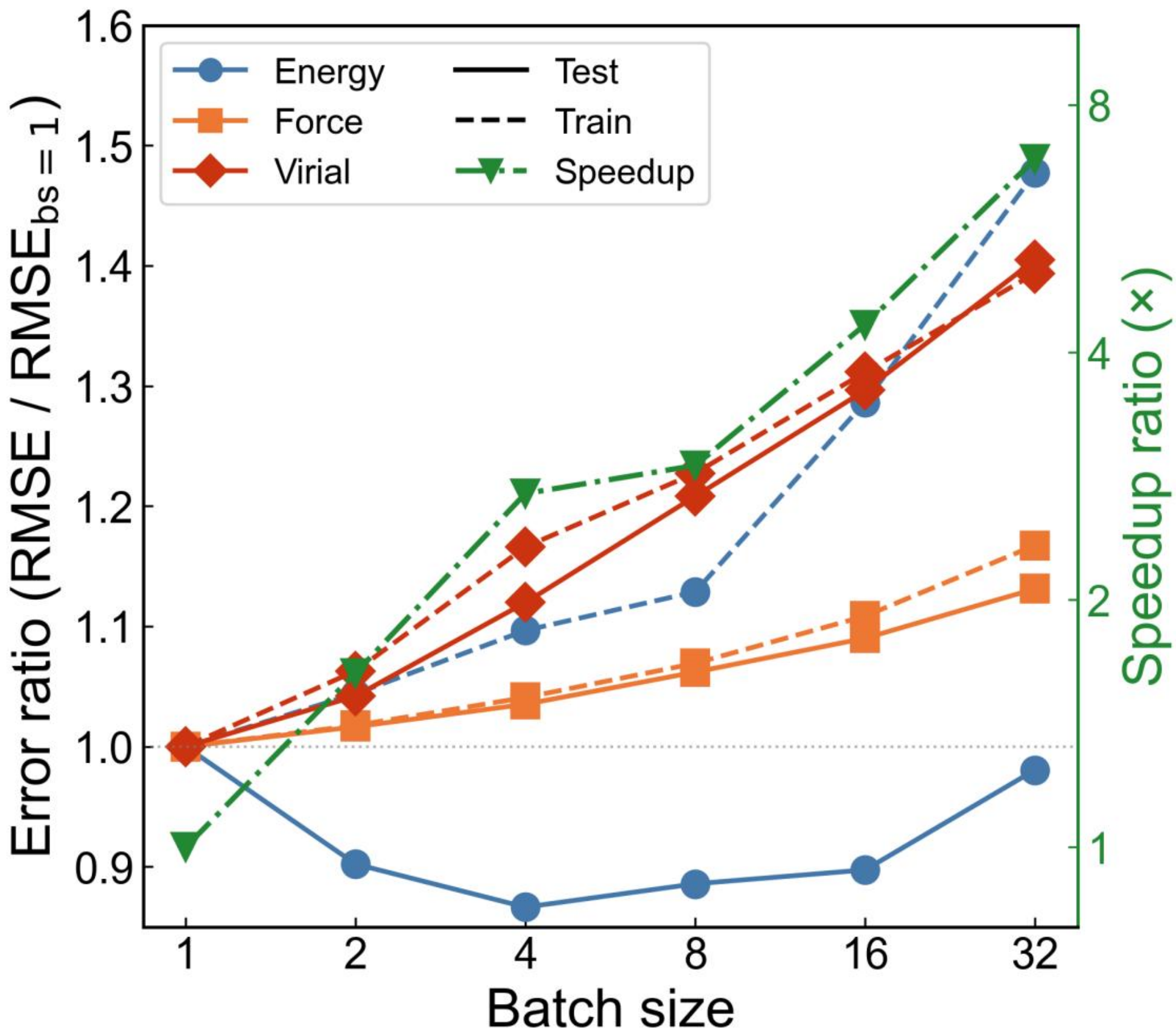

**Figure 10.** Effect of batch size on NEP accuracy and training efficiency.

### 4.2. Training performance

The MTP and NEP models discussed in **Section 4.1** differ in model complexity: the MTP model contains 526 trainable parameters, whereas the NEP model contains 7.9k trainable parameters and uses a larger radial cutoff. Therefore, both the accuracy and training-time comparisons reported here reflect the behavior of these specific model configurations, rather than an intrinsic advantage of one potential form over the other. In this section, we compare the training performance of the same two models, as summarized in **Table 3**.

**Table 3.** Training time (unit: hour) for MTP (in **Section 4.1.1**) and NEP (in **Section 4.1.2**) on the Ge-Sb-Te dataset.

| Model \ Batch size | 1 | 2 | 4 | 8 | 16 | 32 |
|---|---|---|---|---|---|---|
| MTP | 5.93 | 3.40 | 1.90 | 1.40 | 0.90 | 0.75 |
| NEP | 8.93 | 5.48 | 3.32 | 3.07 | 2.07 | 1.30 |

### 4.3. Inference performance in MD simulation

**Figure 11** compares the MD simulation throughput of the MTP and NEP models across six system sizes ranging from 189 to 3.4 million atoms. At 189 atoms, both models deliver nearly identical performance (MTP: 16.6 katom-step/s, NEP: 15.3 katom-step/s). The divergence becomes pronounced as the system size increases. MTP saturates early: its throughput rises to 269.7 katom-step/s at 6804 atoms and reaches a peak of only 355.3 katom-step/s at 3.4 million atoms, representing a modest 1.3-fold improvement over the 6804-atom value. NEP, by contrast, scales far more aggressively, attaining 807.9 katom-step/s already at 55 566 atoms and ultimately peaking at 923.5 katom-step/s at the largest system size. The peak throughput of NEP is approximately 2.6 times that of MTP. This trend differs from the training-time comparison in **Section 4.2**, where MTP exhibits shorter training times than NEP owing to its smaller number of trainable parameters. The slower MTP inference in **Figure 11** is mainly associated with the current CUDA implementation of the MTP inference operator. Specifically, in the MTP kernel, the moment tensor descriptors $M_{\mu,\upsilon}$ require larger local memory than the $S^{i}_{nlm}$ intermediates in NEP, resulting in increased register pressure and higher local-memory traffic during atom-wise CUDA execution. Nsight Compute profiling further shows that both MTP and NEP inference kernels are dominated by L2-cache memory traffic rather than arithmetic operations, with a more pronounced bottleneck in the MTP. Further optimization of both the MTP and NEP inference kernels will be pursued in future work.

All performance results reported in this work were obtained using a single GPU. Although AI2Pot already supports MPI-based multi-GPU execution, its parallel performance will be evaluated and reported in future work.

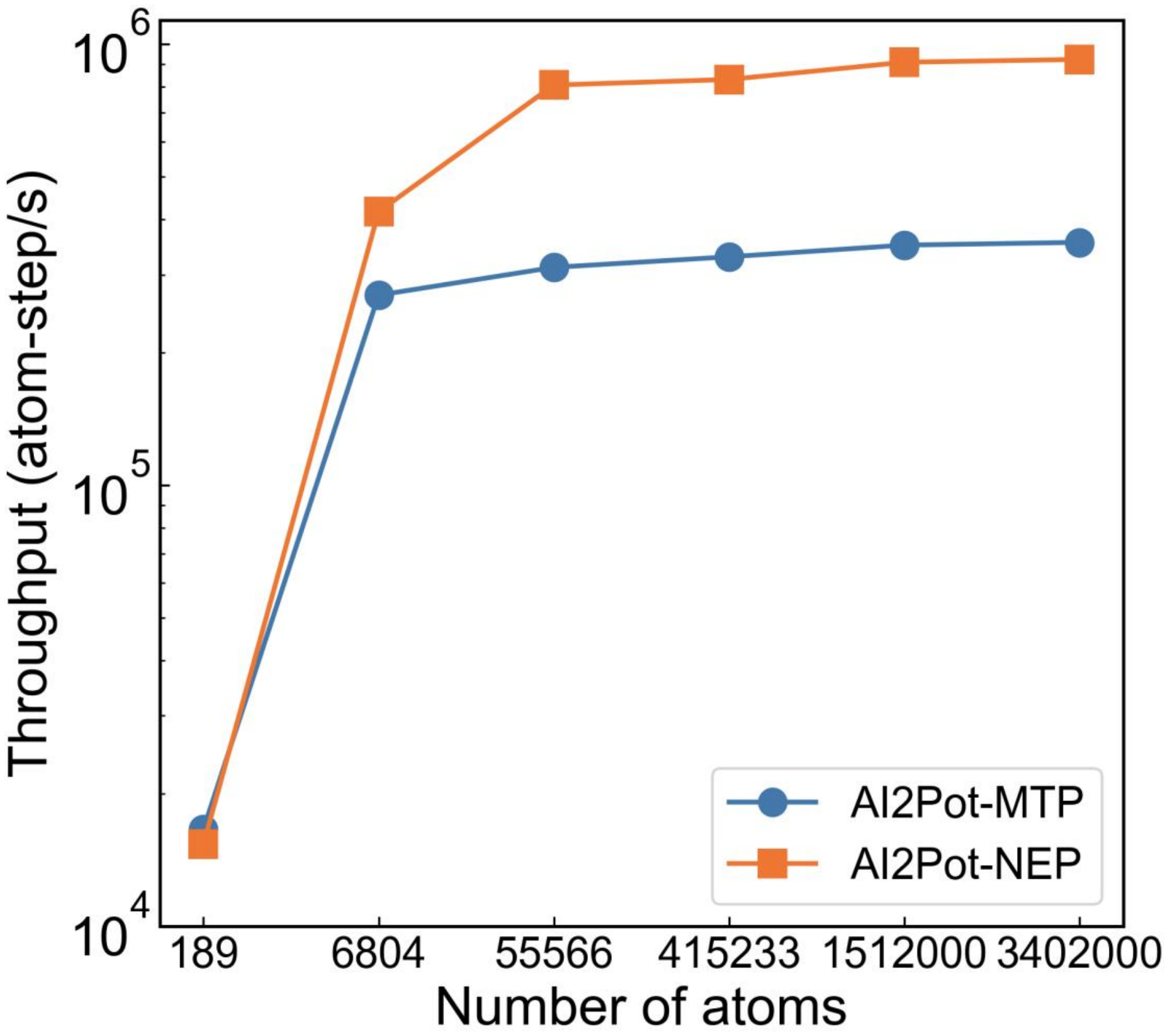

**Figure 11.** MD simulation throughput (katom-step/s) of the MTP and NEP models as a function of the number of atoms.

## 5. Conclusions

In this work, we presented AI2Pot, a scalable and unified machine-learning interatomic potential (MLIP) framework that integrates high-performance atomistic computing within PyTorch-based ecosystem. By re-optimizing C++/CUDA atom-wise operators and integrating them with a PyTorch and PyTorch Lightning frontend, AI2Pot enables performance-critical MLIP operators to be executed efficiently while retaining the flexibility of modern machine-learning workflows.

Crucially, the framework establishes a unified computational backend for both Moment Tensor Potential (MTP) and NeuroEvolution Potential (NEP), enabling different MLIP architectures to be trained and deployed through a consistent software interface. The capability of AI2Pot was demonstrated using a Ge-Sb-Te dataset containing crystalline, liquid, and amorphous configurations. With the default training configurations generated by AI2Pot-cli, both MTP and NEP models achieved reasonable predictive accuracy. The effect of batch size on training accuracy and efficiency is further evaluated. Increasing the batch size substantially reduces the training time for both MTP and NEP, but may also reduce accuracy.

Furthermore, large-scale molecular dynamics simulations via LAMMPS interface demonstrated the scalability and computational efficiency of AI2Pot. AI2Pot models deployed in LAMMPS enable simulations of atomic systems containing millions of atoms on a single GPU, with peak speeds of 355.3 and 923.5 katom-steps/s for MTP and NEP, respectively. Together with AI2Pot-cli, the framework provides a convenient end-to-end route from first-principles datasets to production-oriented MLIP training and large-scale molecular dynamics deployment. Ultimately, AI2Pot not only provides a highly efficient tool for materials scientists but also establishes a unified computational framework that bridges high-performance atomistic simulation and modern machine-learning ecosystems, enabling flexible and efficient MLIP applications.

## Code Availability

AI2Pot and the associated command-line tool AI2Pot-cli are open-source and publicly available under the GNU General Public License v3.0 at https://github.com/lhycms/AI2Pot and https://github.com/lhycms/AI2Pot-cli, respectively. The results reported in this work were obtained using AI2Pot v1.0.0 and AI2Pot-cli v1.0.0.

## Data Availability

The data supporting the findings of this paper are available from the corresponding authors upon reasonable request.

## Acknowledgements

This work is financially supported by the National Key R&D Program of China (2022ZD0117601), and National Natural Science Foundation of China (No.52332005).

## Author Contributions

Hanyu Liu: Conceptualization, Methodology, Software, Validation, Investigation, Data curation, Formal analysis, Visualization, Writing – original draft. Linggang Zhu: Conceptualization, Methodology, Investigation, Formal analysis, Supervision, Funding acquisition, Writing – review & editing. Xuanguang Zhang: Data curation. Ning Yang: Data curation. Jian Zhou: Supervision, Writing – review & editing. Zhimei Sun: Supervision, Funding acquisition, Writing – review & editing.

## Conflict of Interest Statement

The authors declare no conflict of interest.